\documentclass[10pt]{article}
\usepackage{arxiv}
\usepackage[utf8]{inputenc}
\usepackage[T1]{fontenc}
\usepackage{authblk}
\usepackage{graphicx}
\usepackage{amsmath}
\usepackage{amssymb}
\usepackage[capitalise]{cleveref}
\usepackage{natbib}
\usepackage{url}
\usepackage{xcolor}
\usepackage{colortbl}
\usepackage{array}
\AtBeginDocument{\setlength{\headheight}{23pt}}
\title{Enhanced Convergence in p-bit Based Simulated Annealing with Partial Deactivation for Large-Scale Combinatorial Optimization Problems}

\author[1]{Naoya Onizawa}
\author[1]{Takahiro Hanyu}
\affil[1]{Research Institute of Electrical Communication, Tohoku University, Sendai, 980-8577, Japan}
\date{}

\begin{document}
	
	\flushbottom
	\maketitle
	\begin{abstract}
		This article critically investigates the limitations of the simulated annealing algorithm using probabilistic bits (pSA) in solving large-scale combinatorial optimization problems.
		The study begins with an in-depth analysis of the pSA process, focusing on the issues resulting from unexpected oscillations among p-bits. 
		These oscillations hinder the energy reduction of the Ising model and thus obstruct the successful execution of pSA in complex tasks.
		Through detailed simulations, we unravel the root cause of this energy stagnation, identifying the feedback mechanism inherent to the pSA operation as the primary contributor to these disruptive oscillations.
		To address this challenge, we propose two novel algorithms, time average pSA (TApSA) and stalled pSA (SpSA).
		These algorithms are designed based on partial deactivation of p-bits and are thoroughly tested using Python simulations on maximum cut benchmarks that are typical combinatorial optimization problems.
	On the 16 benchmarks from 800 to 5,000 nodes, the proposed methods improve the normalized cut value from 0.8\% to 98.4\% on average in comparison with the conventional pSA.
	%
	\end{abstract}
	%
	%
	\thispagestyle{empty}

\section*{Introduction}

	In recent years, a new device model known as the probabilistic bit, or p-bit, has been proposed \cite{IL}.
	Unlike traditional bits which can only exist in a state of 0 or 1, a p-bit can exist in a range of states between 0 and 1, each state has a certain probability of occurring.
	The p-bit is a versatile computational model that can be implemented in software \cite{p-bit_emulation} or emerging probabilistic devices, such as Magnetoresistive Random Access Memory (MRAM)  \cite{p-bit_device}.
	Furthermore, it can be approximated by digital circuits, such as Field-Programmable Gate Arrays (FPGAs) \cite{IL_FPGA,CIL,p-bit_FPGA,p-bit_async_impl}.
	This probabilistic nature makes p-bits a useful tool in solving certain types of problems that require a degree of randomness or uncertainty. 
	The output state of a p-bit is represented as follows:
	\begin{equation}
		\sigma_i(t+1) = {\rm sgn}\Bigl(r_i(t) + {\rm tanh}\bigl(I_i(t+1)\bigr)\Bigr), 
		\label{eqn:pbits}
	\end{equation}
	where $\sigma_i(t+1) \in \{-1,1\}$ is a binary output signal, $I_i(t+1)$ is a real-valued input signal, and $r_i(t) \in \{-1:1\}$ is a random signal.
	%
		
	The utilization of p-bits is notably effective in the development of a specific neural network variant, the Boltzmann machine \cite{Boltzmann1984}. This model is particularly well-adapted for tasks that require invertible logic \cite{CIL_training}, where inputs and outputs can be interchanged. Moreover, it has significant applications in Bayesian inference \cite{p-bit_BI}, parallel tempering \cite{p-bit_PT}, Gibbs sampling \cite{p-bit_gibbs}, and simulated annealing (SA) \cite{p-bit_general}.

	SA is a stochastic optimization technique widely used for addressing combinatorial optimization problems \cite{SA1,SA2}. Its applications span diverse real-world scenarios, including solving the maximum cut (MAX-CUT) problem in network analysis \cite{SA_max-cut}, optimizing communication systems \cite{SA_LDPC}, and enhancing various machine learning algorithms \cite{SA_ML}.

	Combinatorial optimization problems can often be represented by Ising models, which are mathematical representations of networks or graphs. 
	These problems are often categorized as NP-hard \cite{NP-hard}, meaning that the time required to find the optimal solution tends to grow exponentially with the size of the problem, making them computationally challenging to solve.
	The goal of simulated annealing in this context is to minimize the `energy' of the Ising model, where `energy' is a metaphor for the objective or cost function of the optimization problem. 
	The global minimum energy state corresponds to the optimal solution to the combinatorial optimization problem.
	%
	
	%
	Diverse enhancements and adaptations of SA, including parallel tempering and stochastic simulated annealing (SSA), have been devised to improve efficiency in solving combinatorial optimization problems \cite{PT,SSA}. Additionally, hardware implementations of SA have been explored for rapidly addressing large-scale combinatorial optimization challenges \cite{DA,Ising_PT,JETCAS_SSA}.

	More advanced computational methods such as quantum annealing (QA) \cite{QA1,QA2} have been developed, with expectations of faster processing times compared to SA \cite{QA_Google}.
	However, despite its potential, the realization of quantum annealing is currently restricted due to limitations in device performance. Large-scale problems remain challenging to solve using quantum annealing methods \cite{QA_GI,QA_review}.
    %
    
 	In addition to simulated annealing, various other algorithms have been developed for solving Ising models. These include coherent Ising machines \cite{CIM}, simulated bifurcation \cite{SB}, and coupled oscillation networks \cite{SA_oscillation}. 
    
	Simulated annealing that utilizes p-bits (pSA) is grounded in a probabilistic computing paradigm, which enables its implementation on classical computers. This theoretical framework positions pSA as a potential tool for efficiently solving large-scale problems.

	A potential advantage of pSA is its ability to update nodes in parallel, as opposed to the serial updating method of traditional SA.
	This means that multiple nodes in the network can be updated simultaneously rather than one at a time, which could potentially lead to a faster convergence to the global minimum energy state, speeding up the process of finding the optimal solution.
	Preliminary studies have demonstrated that pSA can effectively solve small-scale problems \cite{p-bit_general}. 
	On the other hand, the efficacy of pSA appears to diminish as the scale of the problem increases.
	%
	
	Simulation studies have indicated that as the size of the problems increases, particularly in the cases of graph isomorphism and MAX-CUT problems, the effectiveness of finding solutions using pSA  significantly diminishes \cite{SSA}.

	One of the challenges is that the energy of the Ising model  does not decrease as expected and remains high, indicating that pSA struggles to find optimal or near-optimal solutions for these larger problems.

	In pSA, the exact reasons behind this limitation remain unclear and are yet to be understood.

	The initial part of this article will involve a detailed analysis of the issues encountered with pSA. 
	This will involve using simulation techniques to study the behavior of the p-bits in detail, with the aim to identify the root cause of the aforementioned problems.
	The analysis identifies that the failure to lower the energy of the Ising model, which is an issue encountered during the optimization process, stems from oscillations occurring among the p-bits. 
	This issue arises because pSA operates as a feedback system, where the output at one stage becomes the input for the next stage, causing the oscillations to occur and impede the reduction of energy.
	Based on the insights gained from the analysis, two new pSA algorithms are introduced: time average pSA (TApSA) and stalled pSA (SpSA). 
	These algorithms aim to counteract the oscillations based on partial deactivation of p-bits, thereby overcoming the main issue identified with the current pSA process.
	These newly proposed algorithms are then put to the test through simulations conducted using Python. 
	The simulations are applied to solve maximum cut (MAX-CUT) problems \cite{max-cut}, which are typical examples of combinatorial optimization problems.
	The simulation results demonstrate that the newly proposed pSA algorithms significantly outperform both the conventional pSA algorithm and traditional SA algorithms
	This implies that these new algorithms may provide a more effective method for solving combinatorial optimization problems, especially for larger problem sizes where traditional methods struggle.

	\section*{Methods}
	
    \subsection*{Problem identification of pSA}

	\begin{figure}[t]
		\centering
		\includegraphics[width=0.8\linewidth]{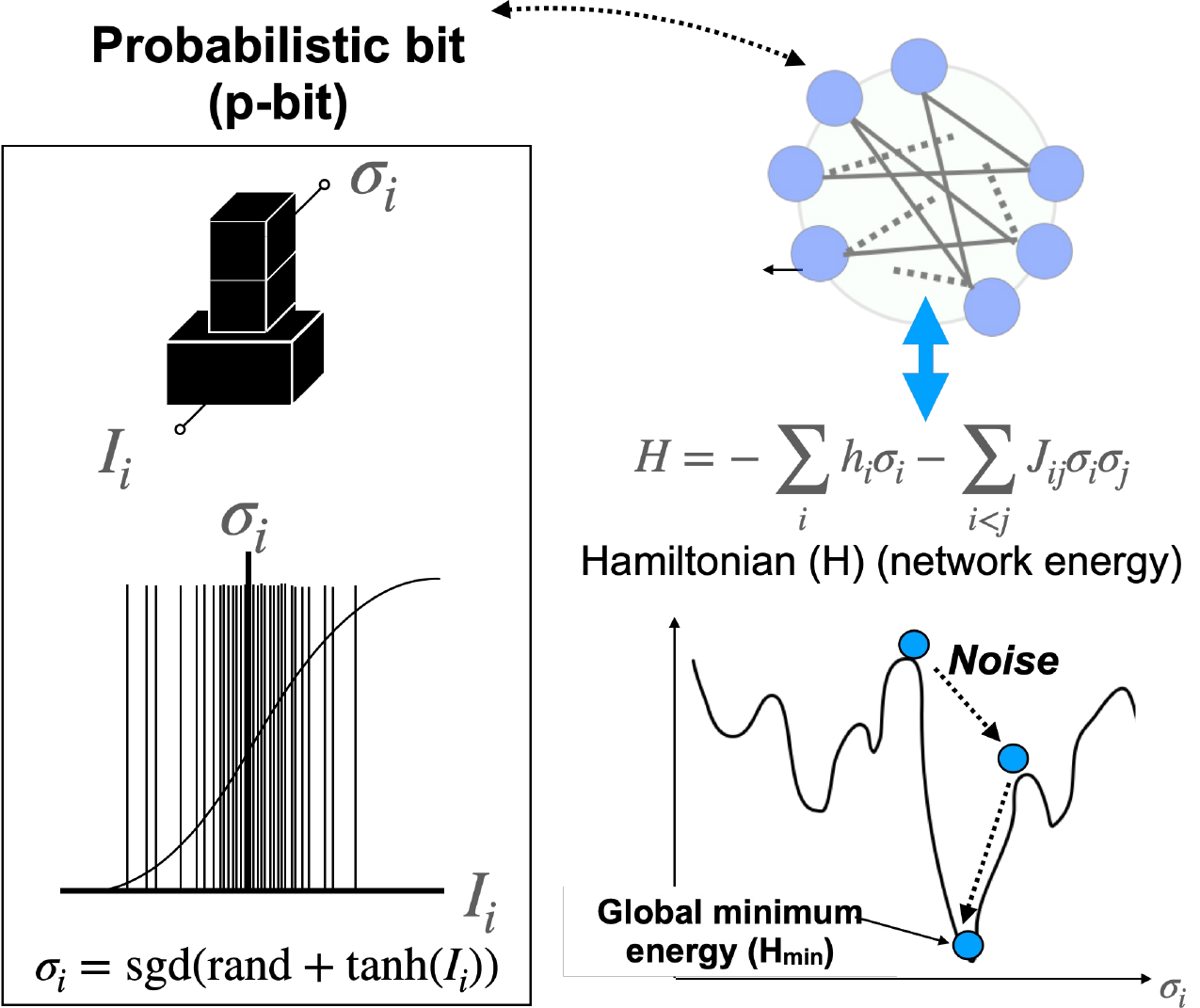}
		\caption{Simulated annealing  based on p-bit (pSA). p-bits (left) probabilistically operates based on \cref{eqn:pbits}. A combinatorial optimization problem  is represented by an Ising model that corresponds to an energy (Hamiltonian). Based on an Ising model, each p-bit is biased with $h$ and is connected with other p-bits with weights $J$ (right top). During the simulated annealing process, an pseudo inverse temperature $I_0$ is gradually increased to reach the global minimum energy ($H_{min}$). pSA attempts to lower the energy of the Ising model by changing the p-bit states $\sigma_i$. If the energy reaches the global minimum energy  $\sigma_i$ are  a solution of the combinatorial optimization problem (right bottom).}
		\label{fig:pSA}
	\end{figure}
	
	A combinatorial optimization problem is represented using an Ising model that represents an energy.
	The energy is represented by Hamiltonian that is defined as follows:
	\begin{equation}
		H(\sigma) = - \sum_i h_i\sigma_i - \sum_{i < j} J_{ij}\sigma_i\sigma_j,
		\label{eqn:Ising}
	\end{equation}
	where  $\sigma_i \in \{-1,1\}$ is a binary state, $h$ are biases for p-bits, and $J$ are weights between p-bits.
	Depending on combinatorial optimization problems, different $h$ and $J$ are assigned.
	Simulated annealing attempts to reach the global minimum energy of  \cref{eqn:Ising} by changing the states $\sigma_i$.
	An algorithm of chancing $\sigma_i$ is different depending on SA algorithms \cite{SA1,p-bit_general,Ising_PT,SSA}.

	pSA \cite{p-bit_general} is illustrated in \cref{fig:pSA}.
	In pSA, each p-bit is biased with $h$ and is connected with other p-bits with weights $J$.
	The input of p-bit $I_i(t+1)$ is calculated using the outputs of other p-bits that is defined as follows:
	\begin{equation}
		I_i(t+1) = I_0 \left (h_i+\sum_j J_{ij}\cdot \sigma_j(t) \right),
		\label{eqn:conv}
	\end{equation}
	where $I_0$ is a pseudo inverse temperature used to control the simulated annealing.
    During the simulated annealing process, $I_0$ is gradually increased in attempt to lower the energy of the Ising model.
    When $I_0$ is small, $\sigma_i$ can be easily flipped between `-1' and `+1' to search for many possible solutions of the combinatorial optimization problem.
    When $I_0$ is large, $\sigma_i$ can be stabilized in attempt to reach the global minimum energy.
    $\sigma_i$ can be found as the solution of the combinatorial optimization problem at the global minimum energy.

	\begin{figure}[t]
		\centering
		\includegraphics[width=0.6\linewidth]{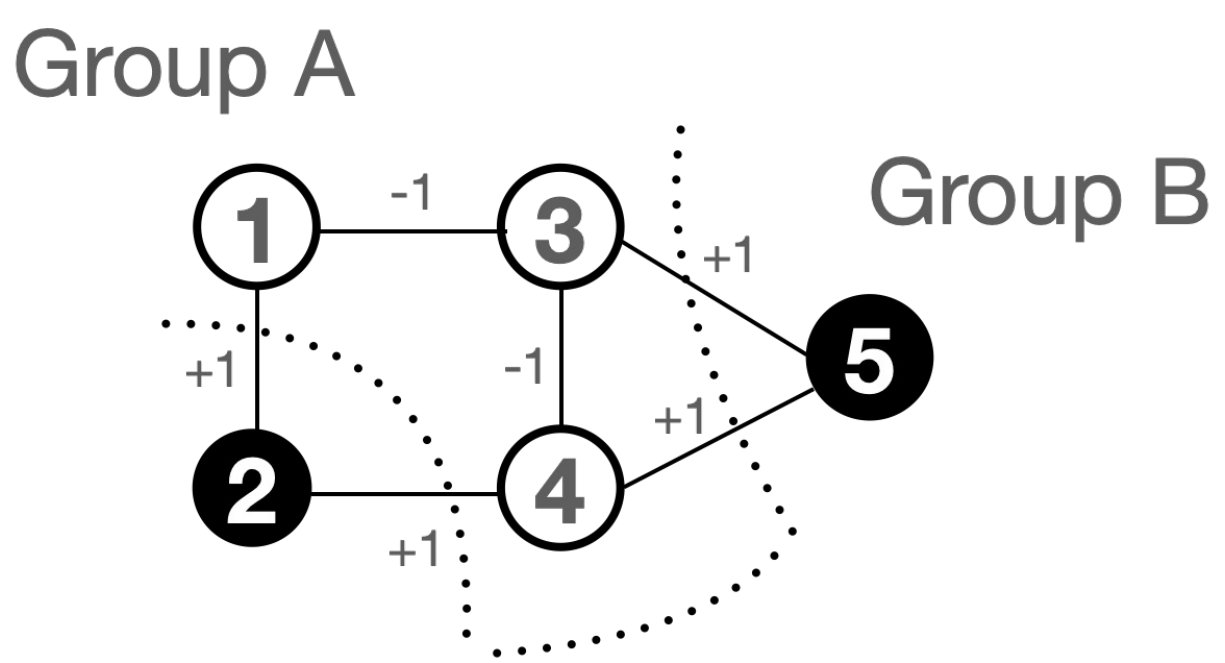}
		\caption{A five-node maximum cut (MAX-CUT) problem with edge weights of $-1$ and $+1$.  MAX-CUT problem  is a typical combinatorial optimization problem. 	The line cuts the edges to divide the graph into two groups while the sum of the edge weights is maximized. The graph is divided into Group A (nodes 1, 3, and 4) and Group B (nodes 2 and 5), with a sum of edge weights equal to 4. }
		\label{fig:maxcut}
	\end{figure}
	
	Let us explain an issue of pSA using a simulated result of  a maximum-cut (MAX-CUT) problem that is a typical combinatorial optimization problem \cite{max-cut} (\cref{fig:maxcut}).
	The MAX-CUT problem aims to partition a graph into two groups in such a way that the sum of the weights of the edges crossing between the two groups is maximized. 
	This process involves `cutting' the graph into two separate sections, hence the term `MAX-CUT'.
	A five-node MAX-CUT problem with edge weights of $-1$ and $+1$ is illustrated (\cref{fig:maxcut}).
	 The black circle illustrates a spin state of `+1', while the white circuit illustrates a spin state of `-1'.
	In the graph, the weight associated with each edge, which can be either -1 or +1, is symbolized by the variable $J$. 
	This variable is crucial in the MAX-CUT problem as it determines the optimal partition of the graph.
	During the simulated annealing process, the spin states are flipped to lower the energy.
	The goal is to find the optimal solution, which corresponds to the minimum energy state.
	After the process, the graph is divided into Group A (nodes 1, 3, and 4) and Group B (nodes 2 and 5), with a sum of edge weights equal to 4.

		\begin{figure}[t]
		\centering
		\includegraphics[width=1.0\linewidth]{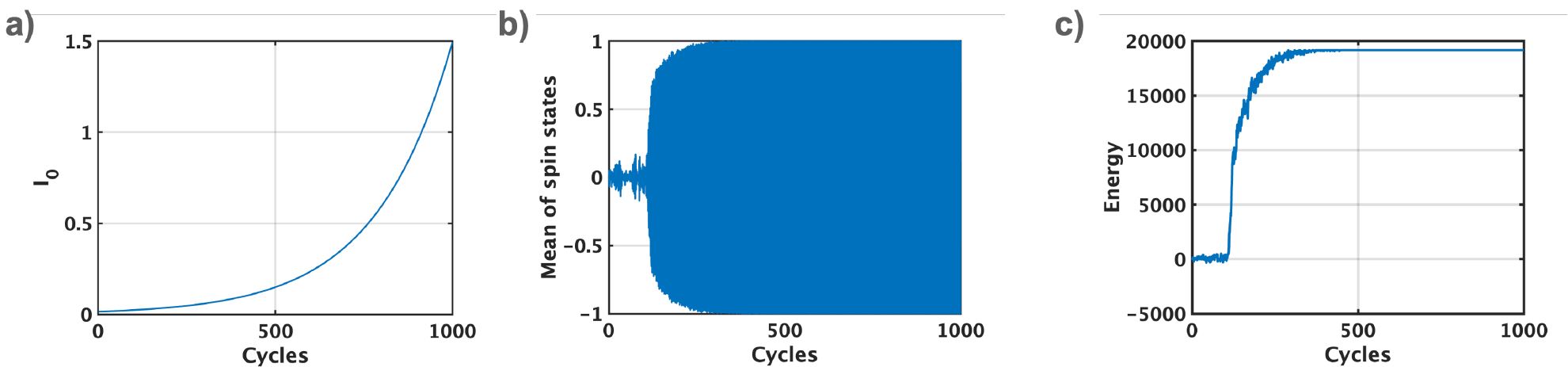}
		\caption{Issue of pSA. pSA is simulated by Python with G1 that is a MAX-CUT problem of the G-set benchmark. During the simulated annealing process, the pseudo inverse temperature $I_0$ is increased to control pSA (a). A mean of all the p-bits states is changed between `-1' and `+1' at every cycle (b). Because of this oscillation, the energy starts to increase after the oscillation, although the energy is expected to be lower to the global minimum energy (c).}
		\label{fig:pSA_issue}
	\end{figure}
	
	pSA with the G1 problem is simulated to identify and understand the current issue of pSA.
	The pSA algorithm is executed using Python 3.11 on Apple M1 Ultra with 128 GB memory.
	G1 is a specific combinatorial optimization challenge called the MAX-CUT problem from the G-set benchmark \cite{G-set}.
	The G1 graph consists of 800 nodes and 19,176 edges that are randomly interconnected.
	To manage the simulated annealing process of pSA, the pseudo inverse temperature $I_0$ is gradually increased over time  from $I_{0min}$ to $I_{0max}$, following the formula $I_0(t+1) = I_0(t)/\beta$, where $\beta$ is 0.995, $I_{0min}$ is 0.0149, and $I_{0max}$=1.49 (\cref{fig:pSA_issue}(a)).
	A method of determining these hyperparameters will be explained in the last subsection.
	During the simulated annealing process, all the p-bit states $\sigma_i$ start to oscillate between `-1' and `+1'.
   The average value of all the p-bit states is changed between `-1' and `+1' at every cycle (\cref{fig:pSA_issue} (b)).
   An unexpected issue arises due to this oscillation: the energy starts to increase rather than decrease towards a global minimum  (\cref{fig:pSA_issue} (c)).
   This suggests that pSA  is not reaching the optimal solution, as we would typically expect the energy to minimize in a successful simulated annealing process.

	\subsection*{Proposed algorithms based on partial deactivation of p-bits}
	
	\subsubsection*{SA based on time average p-bit (TApSA)}
	
	In this article, two new pSA algorithms with nonlinear functions are introduced, which partially deactivates p-bits to mitigate the oscillations.
	The first algorithm is SA based on time average p-bit (TApSA).
	The TApSA algorithm draws its inspiration from stochastic simulated annealing (SSA) \cite{SSA}. 
	SSA approximates the behavior of p-bits using a method called stochastic computing, which is particularly suited for simulated annealing processes.
	Stochastic computing is a computational approach where values are represented as the frequencies (time averages) of 1s in bit streams \cite{stochastic_first,stochastic}. 
	This allows for the efficient operation of time series computations in a way that is hardware efficient in terms of physical area usage \cite{stochastic_book}.
	The use of stochastic computing has been successfully applied in a range of computational applications, such as in low-density parity-check decoders, image processing, digital filters, and deep neural networks \cite{Sldpc1,Simage,SIIR,SDNN}. 
	The SSA approach approximates the tanh function (\cref{eqn:pbits}), a key component of the pSA operation, using a saturated updown counter, which results in an operation that calculates tanh in a time series manner.
	Contrary to pSA, SSA has been shown to be capable of effectively solving large-scale combinatorial optimization problems. 
	This suggests that the new TApSA algorithm, which is inspired by SSA, could potentially address the limitations identified in traditional pSA when dealing with large problem sizes.

   Based on the previously discussed explanations, the time average operation is a key element to solve the issue of pSA.
   The proposed TApSA incorporates a time-average operation into the pSA algorithm. 
   This operation, which is added to \cref{eqn:conv} of pSA, is defined as follows:
\begin{subequations}
	\begin{equation}
		TI_i(t+1) =h_i+\sum_j J_{ij}\cdot \sigma_j(t) ,
		\label{eqn:TApSA1}
	\end{equation}
	\begin{equation}
		  I_i(t+1) =  I_0 \left (\frac{1}{\alpha}\sum_{i=0}^{\alpha-1}TI_i(t+1-i)\right),
		\label{eqn:TApSA2}
	\end{equation}
	\label{eqn:TApSA}
\end{subequations}
where $TI_i(t+1)$ represents a temporary value used for the time averaging operation and the variable $\alpha$ is the size of the time window over which $TI_i(t+1)$ is averaged.
$I_i(t+1)$ is the input for the p-bit, as defined in \cref{eqn:pbits} that is also used in TApSA.
The set of equations in \cref{eqn:TApSA} is responsible for calculating the time average of the p-bit input signal. 
This operation effectively smooths out the signal over a certain time window, which helps to reduce random fluctuations or `noise' in the signal.
Another consequence of this time-averaging operation is that it highlights or emphasizes the lower frequency components of the signal while simultaneously reducing or attenuating the higher frequency components. 

\subsubsection*{SA based on stalled p-bit (SpSA)}

The second algorithm is simulated annealing based on stalled p-bit (SpSA).
SpSA takes inspiration from the sparse random signals used in invertible logic operations \cite{Sparse_random_CIL}.
Invertible logic is an application of p-bits, which permits bidirectional operations of any function. 
Sparse random signals are applied in such a way that they probabilistically halt, or "stall," the addition of random signals to the invertible logic operations, subsequently reducing error rates.
%
%

The SpSA algorithm employs a similar approach, where it probabilistically stalls the behavior of p-bits. 
Specifically, in SpSA, the input of a p-bit, represented as $I_i(t)$, is probabilistically stalled and maintains the same value $I_i(t)$ from the previous time step.
The equation for SpSA as mentioned above is:
%
%
%
\begin{equation}
I_i(t+1) = \begin{cases} 
		I_i(t), & \text{with probability of getting stalled } p \\
	I_0 \left(h_i+\sum_j J_{ij}\cdot \sigma_j(t)\right). & \text{with probability} (1-p) 
\end{cases}
\label{eqn:SpSA}
\end{equation}
%
%
In this equation, the input of the p-bit at time t+1, $I_i(t+1)$, can either be stalled (i.e., be the same as the input at time t, $I_i(t)$), with a probability $p$, or take on a new value with a probability of $(1-p)$.
This approach is a significant deviation from traditional pSA, where \cref{eqn:conv} is replaced by \cref{eqn:SpSA} in SpSA.

   \subsection*{MAX-CUT problems and annealing parameters for evaluation}

   \begin{table}
   	\centering
   	\begin{tabular}{|c|c|c|c|c|c|}
   		\hline
   		\rowcolor{gray!50}
   		Graph  & \# nodes & Structure & Weights ($J$) & \# edges & Best known value \\
   		\hline
   		G1 &800 & random  & {+1} &19176 &   11624\\
   		\hline
   		G6  & 800  & random &{+1, -1} &19176 & 	2178 \\
   		\hline
   		G11  &800  & troidal & {+1, -1}  &1600&564  \\
   		\hline
   		G14  & 800  & planar & {+1} & 4694&	3064\\
   		\hline
   		G18  &  800 & planar & {+1, -1} &4694   & 	992 \\
   		\hline
   		G22 &  2000 & random & {+1} &19990&13359 \\
   		\hline
   		G34  & 2000 & troidal & {+1, -1}  &4000 &1384 \\
   		\hline
   		G38 & 2000 & planar &  {+1} &11779&	7688 \\
   		\hline
   		G39  &  2000 & planar & {+1, -1} &11778 &2408  \\
   		\hline
   		G47&  1000 & random  &{+1} &9990&6657  \\
   		\hline
   		G48 &  3000 & troidal & {+1, -1}&6000 &6000 \\
   		\hline
   		G54 & 1000 & random &  {+1}&5916 &	3852 \\
   		\hline
   		G55 & 5000 & random & {+1}  &12498  & 10299 \\
   		\hline
   		G56  & 5000 & random & {+1, -1}  &12498  &	4017   \\
   		\hline
   		G58 &  5000 & planar & {+1} &29570 &	19293\\
   		\hline
   		K2000 & 2000 & full & {+1, -1}  &1999000  &	33337\\
   		\hline
   	\end{tabular}
   	\caption{
   	Summary of MAX-CUT benchmarks used for evaluating simulated annealing algorithms, including TApSA, SpSA, and traditional SA and pSA. The Gxx graphs are part of the G-set benchmark \cite{G-set}, and K2000 is a fully-connected graph benchmark \cite{K2000}.
   	}
   	\label{tb:graph}
   \end{table}

 \begin{table}
 	\centering
 	\begin{tabular}{|c|c|}
 		\hline
 		\rowcolor{gray!50}
 		Algorithm & Equations \\
 		\hline
 		pSA \cite{p-bit_general} & \cref{eqn:conv,eqn:pbits} \\
 		\hline
 		TApSA (proposed) & \cref{eqn:TApSA,eqn:pbits}\\
 		\hline
 		SpSA (proposed) &  \cref{eqn:SpSA,eqn:pbits} \\
 		\hline
 	\end{tabular}
 	\caption{Summary of the determined hyperparameters for pSA, TApSA, and SpSA. In the simulated annealing process, all the p-bit states are updated using these equations in parallel in attempt to reach the global minimum energy of an Ising model.}
 	\label{tb:eqn}
 \end{table}
 
 \begin{table}
 	\centering
 	\begin{tabular}{|c|c|}
 		\hline
 		\rowcolor{gray!50}
 		Parameter & Value \\
 		\hline
 		$s_i$ & $\sqrt{(n-1)\cdot {\rm Var}(J_{i,:})}$ \\
 		\hline
 		$I_{0min}$ & $\frac{\gamma}{{\rm mean}(s_i)}$ \\
 		\hline
 		$I_{0max}$ & $\frac{\delta}{{\rm mean}(s_i)}$ \\
 		\hline
 		$\beta$ & $\Bigl(\frac{I_{0min}}{I_{0max}}\Bigr)^{(\frac{1}{cycle-1})}$ \\
 		\hline
 	\end{tabular}
 	\caption{Statistically determined hyperparameters for pSA, TApSA, and SpSA. A pseudo inverse temperature is gradually increased over time  from $I_{0min}$ to $I_{0max}$, following the formula $I_0(t+1) = I_0(t)/\beta$ . $\gamma=0.1$ and $\delta=10$ are used in this article.}
 	\label{tb:param}
 \end{table}

   The proposed TApSA and SpSA algorithms are evaluated in MAX-CUT problems using Python 3.11 on Apple M1 Ultra with 128 GB memory.
   Two MAX-CUT benchmarks, namely G-set and K2000, are used for these simulations (\cref{tb:graph}).
    The G-set includes Gxx graphs that vary in node sizes and edge connections \cite{G-set}.
   %
   %
   On the other hand, K2000 represents a fully-connected graph with edge weights of either `-1' or `+1' \cite{K2000}.
   %
   %
   Before the simulated annealing process begins, $J$ of the Ising model is assigned based on the graph weights.
   %

  Performance of the proposed TApSA and SpSA algorithms is compared with the traditional pSA.
   The annealing algorithms are outlined in \cref{tb:eqn}.
   In these simulated annealing algorithms, a crucial factor is the manipulation of the pseudo inverse temperature, a value which has significant implications on annealing to explore the solution space. 
   During the simulated annealing process, the pseudo inverse temperature $I_0$ is gradually increased over time  from the initial value $I_{0min}$ to the maximum value $I_{0max}$, following the formula $I_0(t+1) = I_0(t)/\beta$ (\cref{tb:param}).
   The hyperparameters for the simulated annealing processes, such as $I_{0min}$, $I_{0max}$, and $\beta$, are not arbitrarily selected. 
   Rather, they are determined in accordance with a specific statistical method, which is designed to optimize the performance of the simulated annealing algorithm (SSA) \cite{Hyperparameter_SSA}.
   In addition to these, a traditional SA algorithm, a well-established method for optimization problems, is also implemented for the sake of performance comparison \cite{SA_max-cut}. 
   In the traditional SA algorithm, the annealing temperature $T$ is managed in a slightly different manner: the temperature is gradually decreased at each cycle by a factor of $\Delta_{IT}$, following the equation $T \leftarrow 1/(1/T+\Delta_{IT})$.
    In this experiment, the initial temperature is set to 1, and the final temperature is set to 1/1000.

 All of the simulated annealing algorithms, including TApSA, SpSA, pSA, and traditional SA, are simulated for a total of 1,000 cycles each. 
 This number of cycles allows the annealing system ample opportunity to explore the solution space and converge to a solution.
 Due to the inherent randomness in these probabilistic algorithms, the evaluation is not based on a single trial. Instead, to get a more accurate understanding of their performance, 100 separate trials are executed for each algorithm. 
 The outcomes of these trials are then used to calculate the minimum, average, and maximum cut values of the MAX-CUT problems, providing a comprehensive assessment of the algorithms' performance.

\section*{Results}

\subsection*{Simulation analysis of TApSA}

		\begin{figure}[t]
	\centering
	\includegraphics[width=1.0\linewidth]{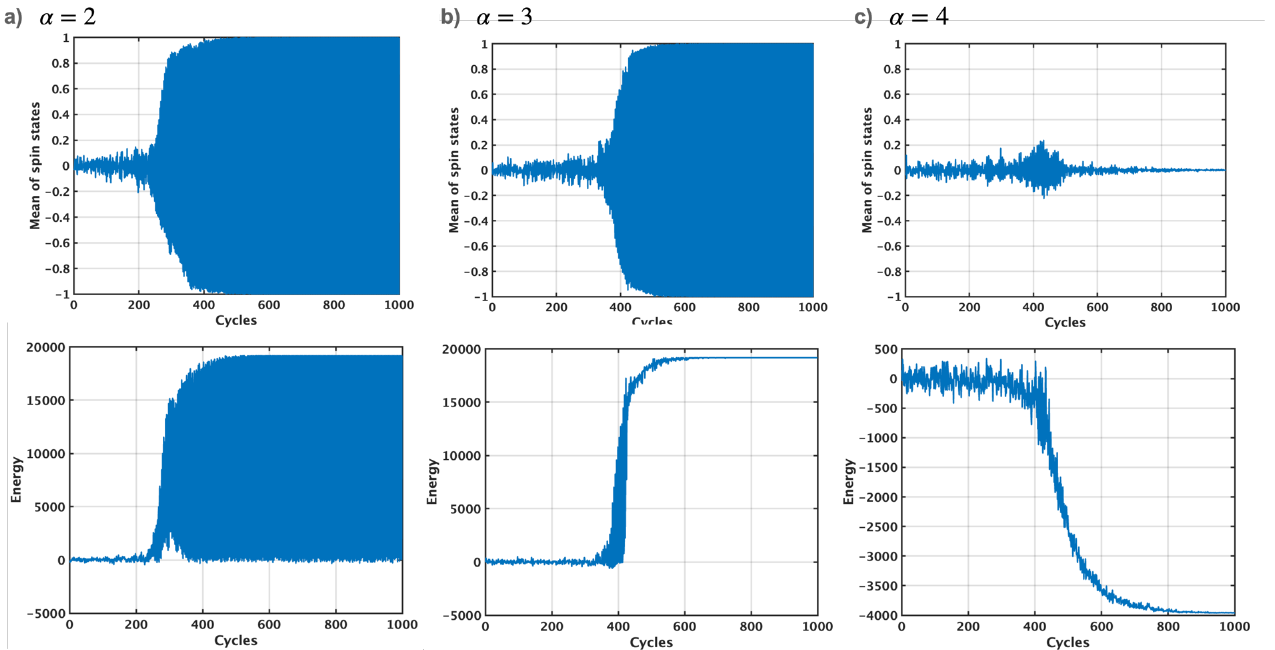}
	\caption{Simulation analysis of TApSA on the G1 graph with different window size $\alpha$. The mean values of all the p-bit states are oscillated between `-1' and `+1' in case of $\alpha$ of two and three (top of a and b), resulting in the energy increase of the Ising model instead of decrease (bottom of a and b). When $\alpha$ is four, no oscillation occurs and hence the energy goes down to the global minimum energy (c). TApSA can solve the oscillation issue of pSA.}
	\label{fig:TApSA}
\end{figure}

The TApSA algorithm is simulated on the G1 graph with varying window size $\alpha$ (\cref{fig:TApSA}).
During the annealing process, the pseudo inverse temperature $I_0$ is gradually increased over time for 1,000 cycles from $I_{0min}=0.0149$ to $I_{0max}=1.49$, following the formula $I_0(t+1) = I_0(t)/0.995$.
$\alpha$ is incrementally increased from two to four to observe the corresponding changes in the behavior of TApSA.
Note that when $\alpha$ equals to one, TApSA operates exactly the same as pSA.
Increasing $\alpha$ can make signals of p-bits smoother.
For $\alpha=2$, the mean of all the p-bit states oscillates between `-1' and `+1'. 
This oscillatory pattern implies that the algorithm alternates between two distinct states throughout its execution. 
This behavior, however, does not lead to the energy of the system decreasing towards the global minimum. 
Instead, it causes an increase in the energy level.
Similarly, when $\alpha$ is increased to three, the oscillation persists, but its start cycle is delayed in comparison to when $\alpha=2$. 
Despite this delayed onset of oscillation, the energy of the Ising model still increases, failing to converge to the global minimum.
However, a significant change in behavior is observed when $\alpha$ is increased to four. 
In this case, no oscillation is observed in the mean of the p-bit states. 
This allows the energy of the Ising model to decrease steadily, eventually reaching the global minimum. 
Thus, for this specific graph and set of conditions, an $\alpha$ value of four appears to facilitate the effective optimization, leading the annealing system towards the global minimum energy state.

	\begin{figure}[t]
	\centering
	\includegraphics[width=0.9\linewidth]{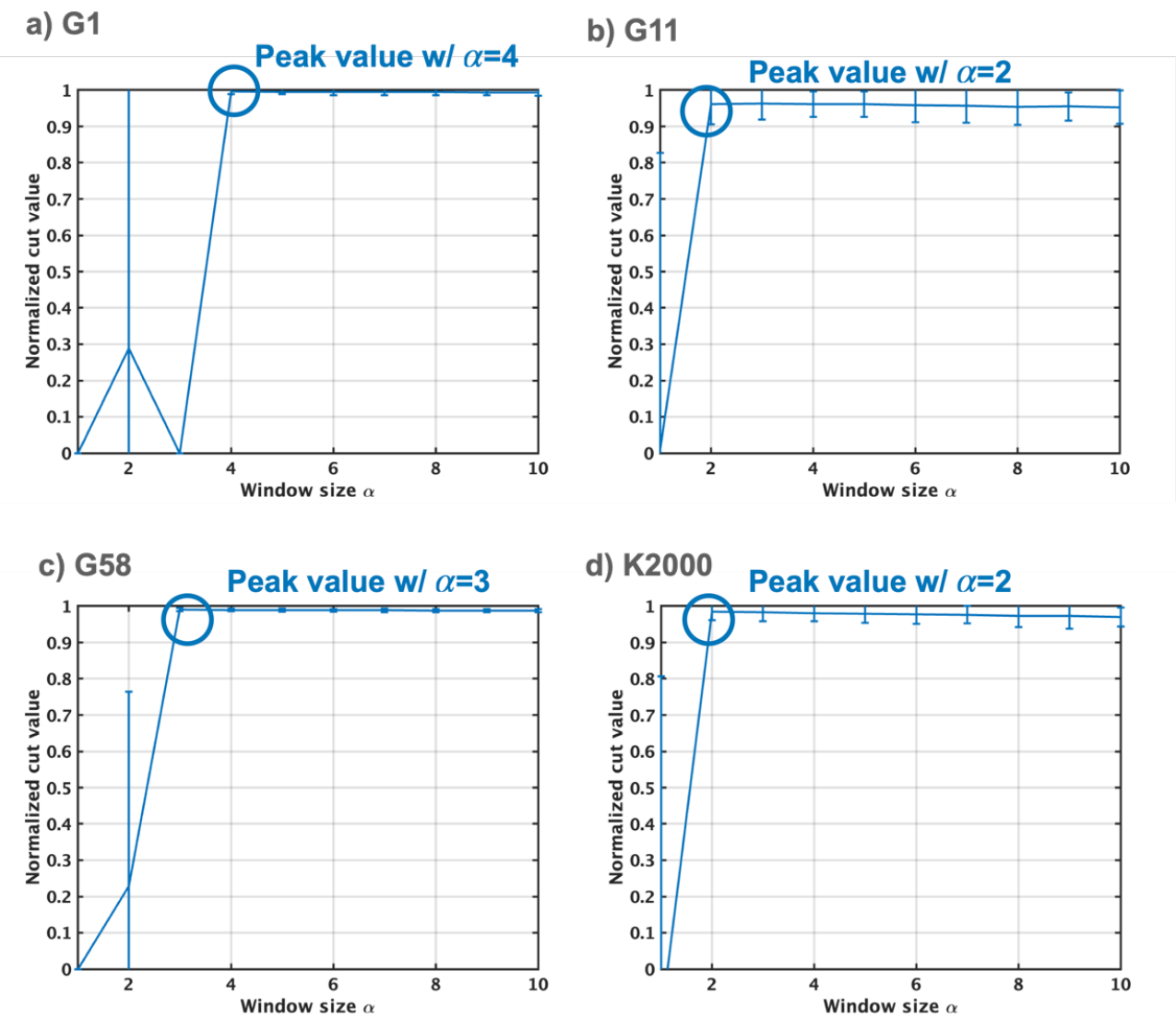}
	\caption{Normalized cut values using the TApSA algorithm on the G1, G11, G58 and K2000 MAX-CUT problems by varying the windows size $\alpha$ from one  to ten. This window size $\alpha$ plays a key role in determining the quality of the solution and can control the behavior of the TApSA algorithm. When $\alpha$ is increased to a specific value, the cut values can be closer to the best-known values because of no oscillation.
	The peak of the normalized mean cut value is obtained with different $\alpha$ depending on the graph.}
	\label{fig:TApSA_performance}
	\vspace{-3mm}
\end{figure}

\begin{table}
	\centering
	\begin{tabular}{|c|c|c|c|c|}
		\hline
		\rowcolor{gray!50}
		Parameter & G1 & G11 & G58 & K2000 \\
		\hline
		$s_i$ & 6.69 & 1.99 & 3.22 & 44.7\\
		\hline
		$I_{0min}$ & 0.0149 & 0.0501 &0.0311 & 0.00224\\
		\hline
		$I_{0max}$ &  1.49 & 5.01 & 3.11 & 0.224\\
		\hline
		$\beta$ &0.995 & 0.995 & 0.995 & 0.995\\
		\hline
	\end{tabular}
	\caption{Summary of hyperparameters to control $I_0$ for pSA, TApSA, and SpSA on the G1, G11, G58, K2000 benchmarks. The number of cycles is 1,000.}
	\label{tb:param_detail}
	\vspace{-3mm}
\end{table}

Next, the TApSA algorithm is simulated to evaluate the normalized cut value on the G1, G11, G58, and K2000 MAX-CUT problems by varying the windows size $\alpha$ from one  to ten (\cref{fig:TApSA_performance}).
By varying the window size, the behavior of the TApSA algorithm and the resulting cut values are influenced. 
To evaluate the performance, the normalized cut values are calculated using the minimum, mean and maximum cut value divided by the best-known value for each benchmark graph.
This normalization process allows for fair comparisons across different graph structures and scales.
The parameters to control the pseudo inverse temperature $I_0$ are summarized in \cref{tb:param_detail}.
These parameters are determined based on \cref{tb:param}.

As $\alpha$ is increased to a specific value, the cut values tend to get closer to the best-known values due to the elimination of oscillation in the algorithm.
However, when $\alpha$ surpasses this optimal value, the normalized cut value starts to decrease slightly as $\alpha$ continues to increase. This suggests that while increasing $\alpha$ can improve performance up to a point, overly large window sizes may have a detrimental effect.
Interestingly, the specific value of $\alpha$ that yields the peak normalized mean cut values varies depending on the graph. 
This indicates that the optimal window size is not universal but instead depends on the particular characteristics of the graph being analyzed.
The source of these differing optimal $\alpha$ values can be traced to the weights present in the respective graphs, as listed in Table \ref{tb:graph}. 
For instance, G1 and G58 only contain weights of `+1', while G11 and K2000 contain both `-1' and `+1' weights. 
This imbalance in weights can lead to strong oscillations in the algorithm. 
Large $\alpha$ values can help to mitigate these oscillations, effectively smoothing the search through the solution space and improving its ability to find the global minimum.

\subsection*{Simulation analysis of SpSA}

\begin{figure}[t]
	\centering
	\includegraphics[width=1.0\linewidth]{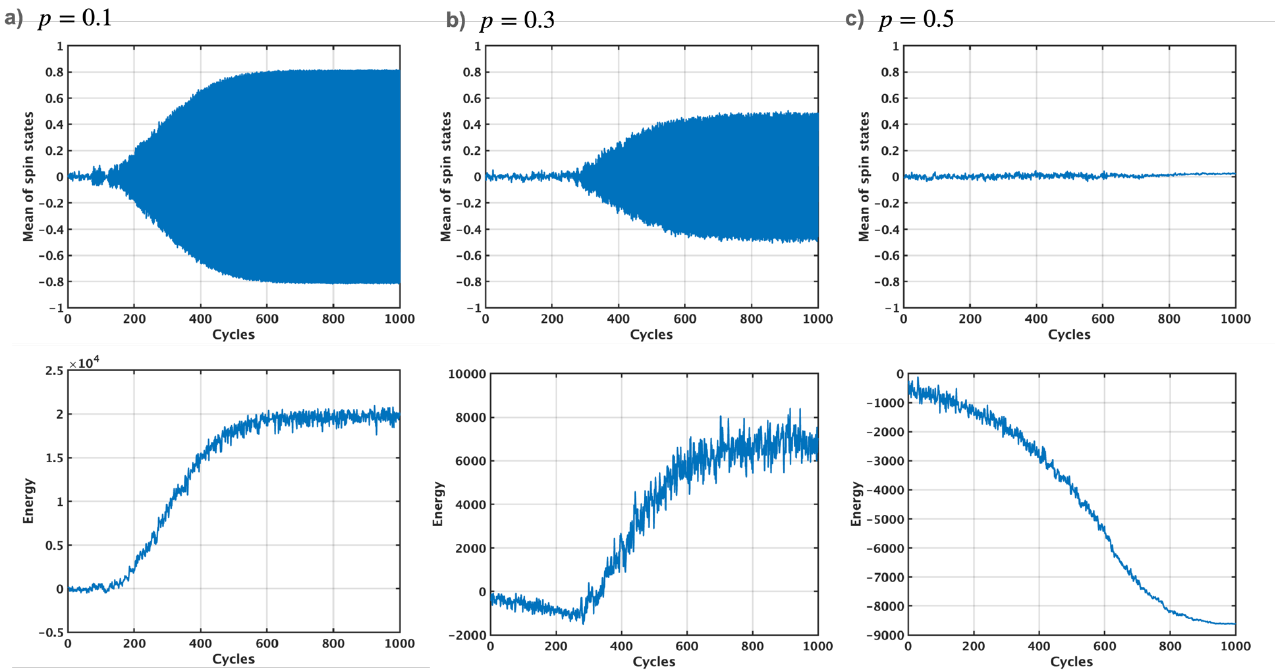}
	\caption{Simulation analysis of SpSA on the G58 graph with different probabilities of getting stalled on p-bits $p$. The mean values of all the p-bit states are oscillated between `-1' and `+1' in case of $p$ of 0.1 and 0.3 (top of a and b), resulting in the increase of energy instead of decrease (bottom of a and b). When $p$ is 0.5, no oscillation occurs and hence the energy goes down to the global minimum energy (c). SpSA can also solve the oscillation issue of pSA.}
	\label{fig:SpSA}
	\vspace{-3mm}
\end{figure}

The G58 graph is used to evaluate the SpSA algorithm, considering varying probabilities of p-bits getting stalled, denoted as $p$ (\cref{fig:SpSA}).
The parameter $p$ is systematically incremented from 0.1 to 0.5 in order to thoroughly understand how changes in $p$ impact the behavior and effectiveness of the SpSA algorithm.
When $p$ equals zero, there is no discernible difference in how SpSA and the pSA algorithm function. 
When $p$ is  0.1, the average of all p-bit states starts to oscillate, alternating between `-1' and `+1'.
This oscillation mirrors the behavior observed in the TApSA algorithm with small $\alpha$, leading to an increase in the energy level of the system, which usually signifies a less optimal solution.
When $p$ rises to 0.3, though the oscillation continues, its magnitude is diminished compared to the scenario where $p=0.1$.
The most significant change in behavior of the SpSA algorithm is observed when $p$ reaches 0.5.
 In this situation, the oscillation of the average p-bit states ceases completely.
  The absence of oscillation allows the energy of the Ising model to decrease steadily. 
  As the energy reduces, the energy moves closer to the most optimal solution, ultimately achieving the global minimum, which represents the best possible solution.

\begin{figure}[t]
	\centering
	\includegraphics[width=0.9\linewidth]{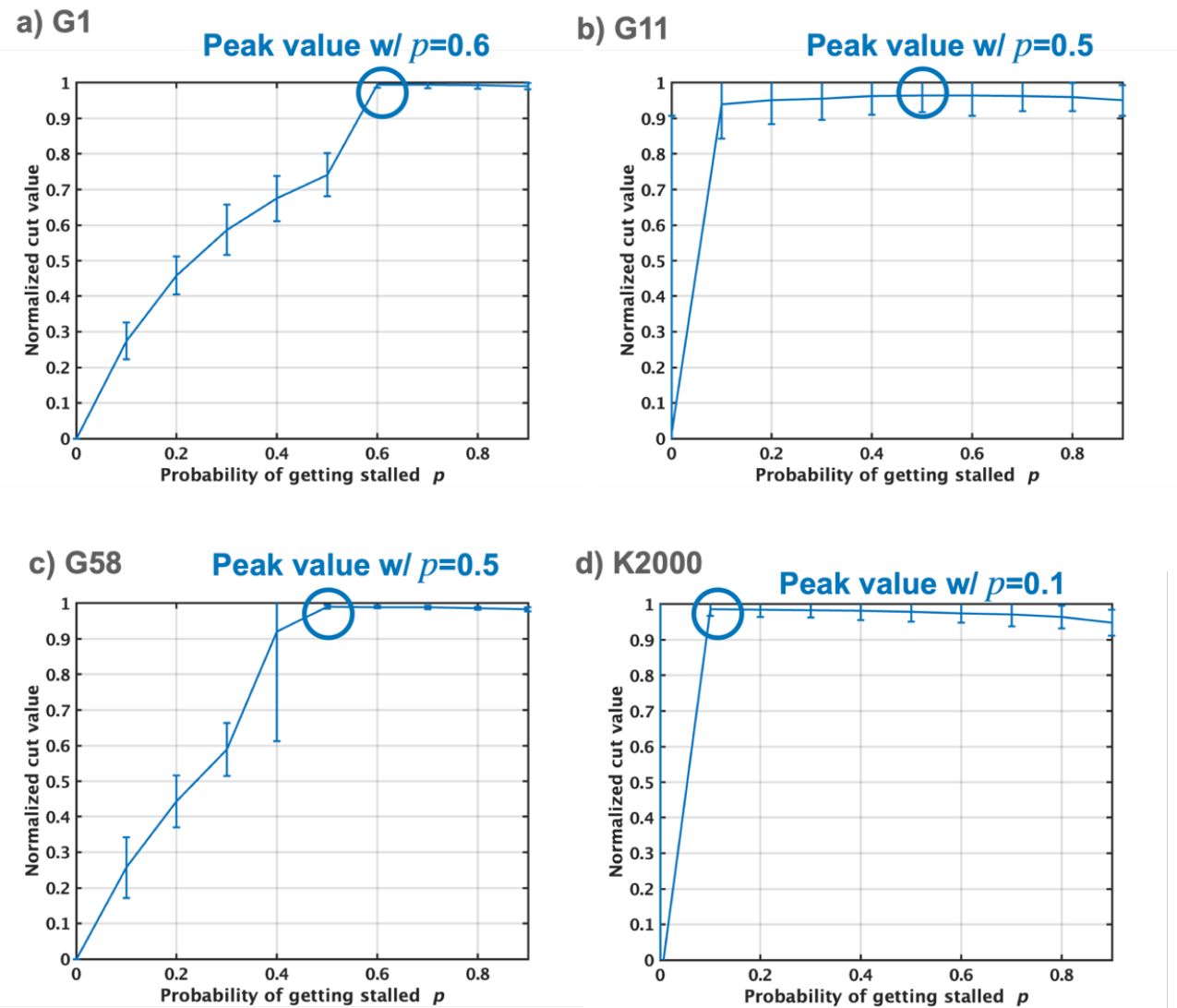}
	\caption{Normalized cut values using the SpSA algorithm on the G1, G11, G58 and K2000 MAX-CUT problems by varying the stalled probability $p$ from 0  to 0.9. When $p$ is increased to a specific value, the cut values can be closer to the best-known values because of no oscillation as well as SpSA.}
	\label{fig:SpSA_performance}
	\vspace{-3mm}
\end{figure}

Next, the SpSA algorithm is simulated to evaluate the normalized cut value on the G1, G11, G58, and K2000 MAX-CUT problems by varying the probability of p-bits getting stalled $p$ from 0  to 0.9 (\cref{fig:SpSA_performance}).
As $p$ is increased to a specific value, the cut values tend to get closer to the best-known values due to the elimination of oscillation as well as TApSA.
When $p$ surpasses this optimal value, the normalized cut value starts to decrease slightly as $p$ continues to increase.
The specific value of $p$ that yields the peak normalized mean cut values varies depending on the graph.

Based on the simulated results of TApSA and SpSA, increasing $\alpha$  exhibits the similar effect to increasing $p$ in terms of eliminating the oscillation, which is crucial for effective optimization.
In both TApSA and SpSA, there appears to be an optimal value of $\alpha$ and $p$, respectively, which yields the best performance in terms of the normalized cut value. 
However, this optimal value is not universal and depends on the specifics of the graph. 
Moreover, surpassing these optimal values can actually lead to a decrease in performance.

\subsection*{Performance comparisons}

	    \begin{figure}[t]
	\centering
	\includegraphics[width=1.0\linewidth]{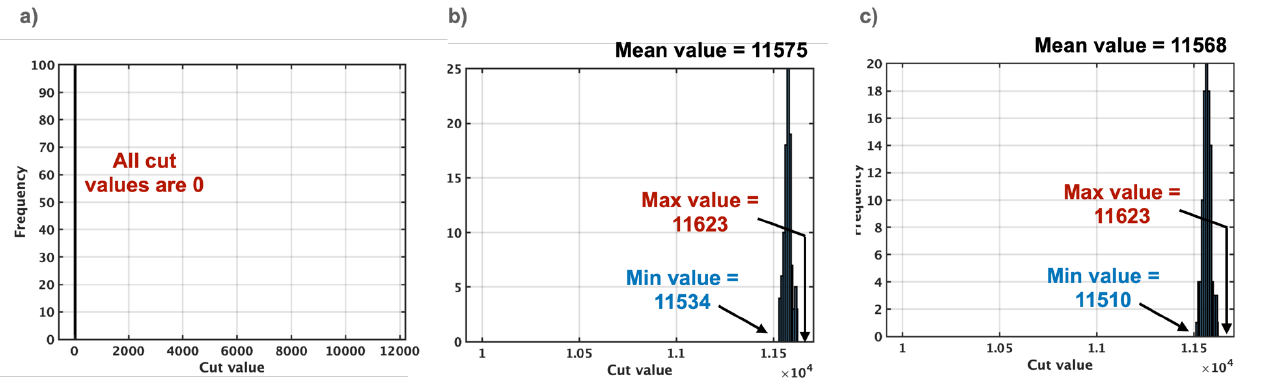}
	\caption{Cut values obtained using simulated annealing  on the G1 graph for 100 trials. The conventional pSA causes all 0 values due to the oscillation (a). In contrast, TApSA with $\alpha=4$ and SpSA with $p=0.6$ can reduce the energy of the Ising model, which results in good cut values that are closer to the best-known values (b and c).}
	\label{fig:hist}
\end{figure}

A comparative analysis of the cut values on the G1 graph are conducted for three different simulated annealing algorithms: pSA, TApSA, and SpSA (\cref{fig:hist}).
To assess the effectiveness of these algorithms in finding cut values, simulations involving 1,000 cycles are performed for each, and these are repeated 100 times. 
The repetition of these simulations leads to the collection of a substantial amount of data, enabling a robust evaluation of the minimum, mean, and maximum cut values.
All cut values using pSA are found to be zero, an intriguing outcome attributable to the oscillation that is observed during the annealing process. 
This suggests that the pSA algorithm is unstable under these conditions, leading to a failure to produce any viable cut values.
%
%
In the case of the TApSA and SpSA algorithms, they are simulated with the most advantageous parameters, namely $\alpha=4$ and $p=0.6$, respectively. 
The choice of these specific values is determined on previous experiments and analysis, which demonstrated superior performance.
%
%
When comparing the results, it is shown that both TApSA and SpSA outperform pSA, achieving near-optimal solutions. 
This is largely due to the elimination of oscillations observed in the pSA algorithm, made possible by the design of the TApSA and SpSA algorithms. 
%

\begin{table}[t]
	\centering
	\begin{tabular}{|c|c|c|c|c|c|c|}
		\hline
		   		\rowcolor{gray!50}
		Graph & SA \cite{SA1} & pSA \cite{p-bit_general} & \multicolumn{2}{c|}{TApSA (proposed)} & \multicolumn{2}{c}{SpSA (proposed)} \\
	     \cline{2-7}
		& Mean cut value & Mean cut value & Mean cut value & Window size  & Mean cut value & Probability of  \\
		& & & &$\alpha$ & & getting stalled $p$\\
		\hline
		G1 & 10757.91 & 0 & 11574.69 & 4 & 11567.89 & 0.6 \\
		G6 & 1270.88 & 173.48 & 2150.49 & 2 & 2151.23 & 0.1 \\
		G11 & 336.72 & 6.18 & 542.7 & 3 & 543.78 & 0.5 \\
		G14 & 2801.84 & 0 & 3035.74 & 3 & 3034.78 & 0.5 \\
		G18 & 591.5 & -49.79 & 968.31 & 2 & 968.94 & 0.1 \\
		G22 & 11161.58 & 0 & 13277.55 & 3 & 13271.27 & 0.5 \\
		G34 & 469.22 & -29.62 & 1331.22 & 2 & 1335.72 & 0.5 \\
		G38 & 6638.96 & 0 & 7617.3 & 3 & 7610.48 & 0.5 \\
		G39 & 854.57 & -489.49 & 2343.52 & 2 & 2349.57 & 0.2 \\
		G47 & 5851.46 & 0 & 6623.31 & 3 & 6618.35 & 0.6 \\
		G48 & 3563.94 & 3057.22 & 5867.16 & 2 & 5897 & 0.1 \\
		G54 & 3479.42 & 0 & 3815.16 & 3 & 3811.77 & 0.5 \\
		G55 & 6968.09 & 0.03 & 10184.66 & 2 & 10193.41 & 0.2 \\
		G56 & 696.99 & -185.22 & 3900.35 & 2 & 3912.14 & 0.1 \\
		G58 & 15787.85 & 0 & 19108.08 & 3 & 19096.28 & 0.5 \\
		K2000 & 11369.62 & -4889.64 & 32812.64 & 2 & 32860.58 & 0.1 \\
		\hline
	\end{tabular}
		\caption{Comparisons of mean cut values in the MAX-CUT benchmarks. On the 16 benchmarks from 800 to 5,000 nodes, the proposed methods improve the normalized cut value from 0.8\% to 98.4\% on average in comparison with the conventional pSA.}
	\label{tb:cut_values}
\end{table}

Furthermore, a comparison is also conducted between the proposed algorithms (TApSA and SpSA) and the traditional SA method, specifically using the G-set and K2000 benchmarks. 
The detailed results of this comparison are summarized in \cref{tb:cut_values}.
The pSA algorithm is found to consistently fail to lower the energy of the Ising model, leading to particular patterns in the mean cut values. 
When the weight values of the graphs are solely `+1', the mean cut values turns out to be 0.
Conversely, when the weight values are either `-1' or `+1', the mean cut values are negative.
On the 16 benchmarks, the normalized mean cut value of pSA is only 0.8\% on average.
The traditional SA algorithm, on the other hand, delivers better cut values compared to pSA. 
However, it is worth noting that these values are still significantly off from the best-known values, where the normalized mean cut value is 44.4\% on average.
Remarkably, the proposed algorithms, TApSA and SpSA, demonstrate substantial superiority over both pSA and traditional SA. 
TApSA and SpSA achieve the normalized mean cut value of 98.3\% and 98.4\% on average, respectively, in all 16 benchmarks used in this study. 
This underlines the effectiveness of these proposed methods and their potential for practical applications in solving similar optimization problems.
%

	
	
	
	
	
	\section*{Discussion}
	
	In this article, we have critically examined the limitations of the simulated annealing using probabilistic bits (pSA) algorithm, specifically with large-scale combinatorial optimization problems such as the maximum cut (MAX-CUT) problem.
	Our detailed analysis has identified disruptive oscillations as the root cause of energy stagnation in the pSA process.
	To mitigate this, we have proposed and rigorously tested two novel algorithms, time average pSA (TApSA) and stalled pSA (SpSA).
	The results suggest significant performance improvements over traditional methods (SA and pSA), highlighting the potential of our proposed algorithms in effectively tackling large-scale optimization tasks.

	Stochastic simulated annealing (SSA) is another p-bit-based simulated annealing that outperforms pSA and SA in several combinatorial optimization problems \cite{SSA}.
	SSA is implemented using digital hardware because the tanh function is approximated using stochastic computing.
	In terms of energy consumption, TApSA and SpSA can reduce energy consumption than SSA because of the nature of their implementation.
	When the p-bit is implemented using an emerging device, the energy consumption can be 10 times smaller than that implemented using a traditional digital circuit \cite{p-bit_device}.
	In addition, the p-bit is a single device, while it can be approximated using several hundreds of transistors in digital implementation, resulting in a more compact hardware implementation.
	In the future, large-scale p-bit-based simulated annealing, TApSA and SpSA could gain in terms of energy and area consumption in comparison with SSA.
	
	\begin{figure}[t]
		\centering
		\includegraphics[width=1.0\linewidth]{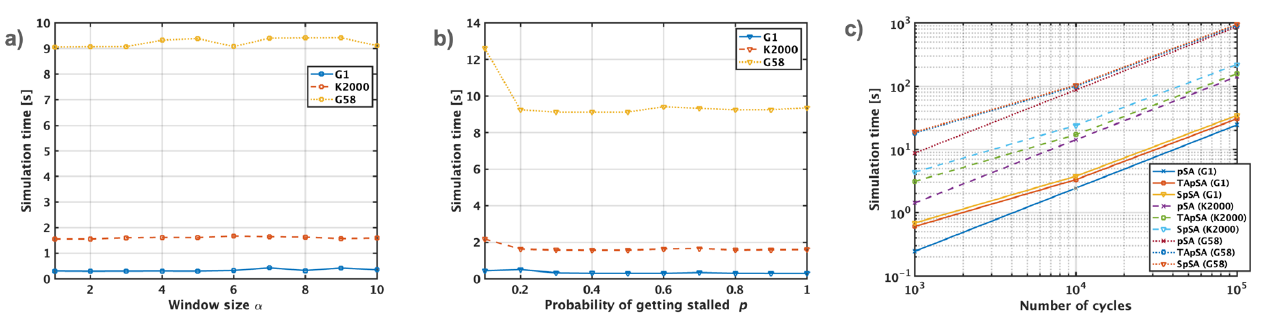}
		\caption{ 
			The simulation duration for TApSA and SpSA was assessed by varying the parameters $\alpha$ and $p$, respectively, across 1,000 cycles (as shown in figures a and b). This evaluation includes the total time spent on the simulations over the number of cycles, incorporating the time required for tuning the parameters $\alpha$ or $p$ (c).
			}
		\label{fig:time}
		\vspace{-3mm}
	\end{figure}

	In terms of computation cost, TApSA and SpSA require extra computation for nonlinear functions in comparison with pSA.

	In \cref{fig:time}(a), the simulation time for TApSA is plotted against the parameter $\alpha$, while \cref{fig:time}(b) illustrates the simulation time for SpSA as a function of $p$, with both scenarios considering 1,000 cycles for solving problems G1, K2000, and G58. 
	It is observed that TApSA's simulation time remains relatively constant regardless of variations in $\alpha$, across the same number of cycles. 
	In contrast, SpSA exhibits longer simulation times at smaller values of $\alpha$, as opposed to shorter times at larger $p$ values. 
	This consistent behavior is further detailed in \cref{fig:time}(c), which shows the total simulation time, inclusive of the duration spent tuning $\alpha$ or $p$. 
	Prior to the annealing process, parameter tuning for TApSA and SpSA involves adjusting $\alpha$ or $p$ over 100 cycles to identify the optimal parameter settings. When compared to pSA, it is noteworthy that the additional computational costs associated with TApSA and SpSA are less significant at higher cycle counts.
    %
	%
	  %
	  %
	  %
	  %
	  
	  %
	From another point of view, it would be preferable to realize new emerging devices that mimic these algorithms for future implementation.
	
	\begin{figure}[t]
		\centering
		\includegraphics[width=1.0\linewidth]{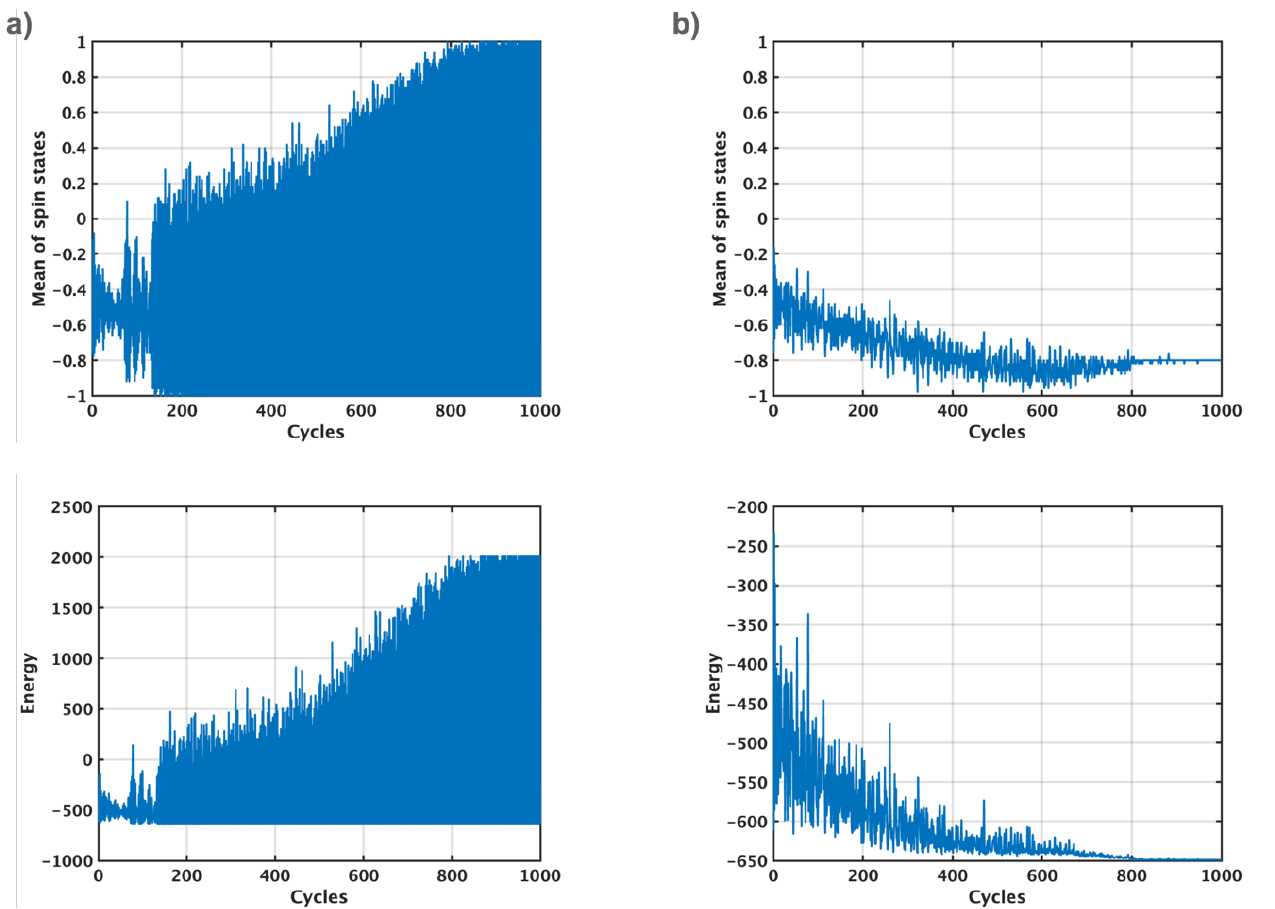}
		\caption{
			In the context of a 100-spin graph isomorphism (GI) problem, which involves determining if two 10-node graphs are isomorphic, distinct performance characteristics of pSA and TApSA with $\alpha=10$ are observed. In case (a), pSA struggles to minimize energy due to oscillations in p-bit states, a challenge similarly noted in the MAX-CUT problem. Conversely, TApSA effectively addresses this oscillation issue, as shown in case (b), leading to a significant reduction in energy that can reach the global minimum. 
			}
		\label{fig:GI}
	\end{figure}

	To determine if pSA induces oscillation in optimization problems other than MAX-CUT, a 100-spin graph isomorphism (GI) problem is employed. 
	This problem involves ascertaining the isomorphism of two 10-node graphs. 
	GI problems are inherently more complex than MAX-CUT problems, primarily due to the presence of non-zero values of $h$ in GI, in contrast to the all-zero values of $h$ in MAX-CUT.
	 Previous literature \cite{SSA} noted that pSA failed to converge in GI problems with more than 25 spins, though it did not explicitly address the oscillation of p-bit states. 
	 To further explore this issue, both pSA and TApSA are applied to the 100-spin GI problem over 1,000 cycles as shown in \cref{fig:GI}.
	 The result shows that pSA, similar to its performance in MAX-CUT problems, is unable to reduce energy due to p-bit state oscillation. 
	 On the other hand, TApSA effectively resolves this oscillation issue, achieving a reduction in energy to the global minimum. 
	 These findings suggest that while TApSA shows promise in addressing oscillation issues in various algorithms, a more detailed analysis in other contexts is planned for future work
	%
	%
	%
	%
	%
	%
	%
	%

	\begin{table}
		\centering
		\begin{tabular}{|c||c|c|c|}
			\hline
			\rowcolor{gray!50}
			Benchmark & GPU \cite{Ising_GPU} & SimCIM \cite{Ising_parallel} & TApSA \\
			\hline
			G22 & N/A& 99.6\%& 99.4\%  \\
		    G39 & N/A & 97.9\%& 97.3\% \\
		    G47 & 99.4\% &N/A &99.5\% \\
		    G54 & 97.7\% &N/A &99.0\% \\
		    K2000 &N/A & 99.6\% & 98.4\% \\
		    \hline
		\end{tabular}
		\caption{ 
			Comparisons of normalized mean cut values obtained in this research with those reported in related works \cite{Ising_GPU,Ising_parallel}. In this comparison, both the GPU-based approach and TApSA are configured to perform 1,000 annealing steps. In contrast, the SimCIM approach, as referenced in the literature, requires a significantly higher number of annealing steps, totaling 50,000. This disparity in the number of annealing steps highlights the efficiency differences among these methods.
			}
		\label{tb:comp}
	\end{table}

	The performance of the newly proposed TApSA algorithm is benchmarked against other notable methods, specifically the GPU-based asynchronous parallel algorithm \cite{Ising_GPU} and the coherent Ising machine (CIM) \cite{CIM}, with details presented in \cref{tb:comp}. 
	This comparison uses normalized mean cut values, calculated by dividing the mean cut value by the best-known value for each problem. 
	The GPU-based approach evaluates mean cut values over 1,000 annealing steps. 
	Conversely, the performance data for CIM, sourced from the literature \cite{Ising_parallel}, is based on simulations using SimCIM \cite{SimCIM} for a substantially longer duration of 50,000 annealing steps. 
	In comparison with the GPU-based method, TApSA achieves comparable normalized mean cut values. 
	While TApSA's performance is marginally lower than CIM, it shows potential for improvement with an increased number of annealing steps.
	%
	%
	%
	%
	%
	%

\begin{table}
	\centering
	\begin{tabular}{|c||c|c|c|}
		\hline
		\rowcolor{gray!50}
		$r_i(t)$ & pSA & TApSA & SpSA \\
		\hline
		Uniform &0.828\%&98.3\% &	98.4\% \\
		Poisson  & 3.94\%	&97.9\% &	97.9\% \\
		\hline
	\end{tabular}
	\caption{ Normalized mean cut values on average for all 16 benchmarks with different random signals.}
	\label{tb:rand}
\end{table}

  Initially, p-bits are modeled with uniform random signals, where $r_i(t) \in \{-1:1\}$, as outlined in \cref{eqn:pbits}. 
  To explore the impact of random signals on performance, this study introduces random signals derived from a Poisson distribution for the p-bits. 
  Specifically, the random signal $r_i(t) = 1/\lambda \cdot X-1$ is generated according to the Poisson probability formula $P(X=k) = e^{-\lambda} \lambda^k / k!$, with $\lambda$ set to 10. 
  A comparison of normalized mean cut values, using both uniform random and Poisson distribution-based signals, is presented in \cref{tb:rand}. 
  The results indicate that the type of random signals has a negligible effect on the performance of all three algorithms under study.
  %
  %
  %
  %
  %

  P-bits have found application in various domains, one of which includes Gibbs sampling \cite{p-bit_gibbs}. 
  A key distinction between Gibbs sampling and pSA lies in the approach to node updates: Gibbs sampling typically operates serially, while pSA updates nodes in parallel. 
  It is important to note that although extended versions of Gibbs sampling, such as chromatic Gibbs sampling, have the capability to operate in parallel \cite{p-bit_async_impl}, the scope of their applications is relatively limited.
  Simulation results have shown that pSA faces an oscillation issue due to its parallel update mechanism, a problem which the proposed algorithms aim to address. 
  However, applying techniques based on TApSA and SpSA to Gibbs sampling, which inherently operates serially, presents a significant challenge. 
  Exploring a Gibbs sampling method that incorporates the proposed techniques is an intriguing direction for future research.
  %
  %
  %
  %
  %
  %
  %

	In conclusion, our research has broadened the understanding of the pSA process and has led to the development of more effective algorithms for complex optimization tasks.
	The proposed TApSA and SpSA algorithms offer promising avenues for overcoming the limitations of the traditional pSA approach and could be crucial for future progress in combinatorial optimization.

	\section*{Data availability}
	All data generated or analyzed during this study are included in this published article. The Python codes are available at  https://github.com/nonizawa/pSA.
	
	\bibliographystyle{unsrtnat}
	\bibliography{SR}

	\section*{Acknowledgments}
	
	This work was supported in part by JST CREST Grant Number JPMJCR19K3, and JSPS KAKENHI Grant Number JP21H03404.
	
	\section*{Author contributions statement}
	N. O. conducted and analyzed the experiments. T. H. discussed the experiment. All authors reviewed the manuscript. 
	
	\section*{Additional information}
	Competing financial interests: The authors declare no competing financial interests.
	
	
	
	
\end{document}